\documentclass[a4paper,11pt]{article}
\usepackage{pos}

\usepackage{subcaption}

\title{Dimuon production in DIS with charm-mass effects}

\author{I. Helenius}
\author{H. Paukkunen}
\author*{S. Yrjänheikki}

\affiliation{University of Jyväskylä, Department of Physics, P.O. Box 35, 40014 University of Jyväskylä, Finland}

\affiliation{Helsinki Institute of Physics, P.O. Box 64, 00014 University of Helsinki, Finland}

\emailAdd{ilkka.m.helenius@jyu.fi}
\emailAdd{hannu.t.paukkunen@jyu.fi}
\emailAdd{sami.a.yrjanheikki@jyu.fi}

\abstract{
Dimuon production in neutrino-nucleus collisions, an important constraint of strangeness in global parton distribution function analyses, is typically calculated by assuming it to be proportional to inclusive charm production. This approach breaks down beyond fixed-flavor leading-order calculations. In our previous work, we introduced an alternative approach based on semi-inclusive charmed-hadron production to compute dimuon production directly. We now present an extension to this work, where we compute the semi-inclusive hadron production in the SACOT-$\chi$ general-mass variable-flavor number scheme to take all charm-mass effects consistently into account. The results are in line with our expectations, with the dynamical charm-mass effects modifying our previous approximative calculation by up to $20 \, \%$ at small values of $Q^2$.
}

\FullConference{XXXII International Workshop on Deep Inelastic Scattering and Related Subjects (DIS2025)\\
24-28 March, 2025\\
Cape Town, South Africa\\}

\renewcommand{\hookAfterAbstract}{%
	\par\bigskip
	\textsc{ArXiv ePrint}:
	\href{https://arxiv.org/abs/2506.09492}{2506.09492}
}

\newcommand{\renorm}{\mu_{\text{ren}}}
\newcommand{\fact}{\mu_{\text{fact}}}
\newcommand{\frag}{\mu_{\text{frag}}}
\newcommand{\dd}{\mathrm{d}}

\begin{document}
\maketitle

\section{Introduction}

One of the more poorly constrained distributions in global analyses of parton distribution functions (PDFs) is the strange-quark distribution \cite{Hou:2019efy,Bailey:2020ooq,NNPDF:2021njg,Eskola:2021nhw,Duwentaster:2022kpv,AbdulKhalek:2022fyi}. Charm production in charged-current deep inelastic scattering (DIS) remains an important constraint of strangeness to this day \cite{CCFR:1994ikl,NuTeV:2001dfo,NuTeV:2007uwm,CHORUS:2008vjb,NOMAD:2013hbk}. As charm production is not directly observable, dimuon production in neutrino-nucleus collisions is used as the constraint instead. However, dimuon production is still usually calculated by multiplying the inclusive charm production cross section with corrective factors. This approach breaks down beyond the fixed-order leading-order (LO) calculation. We take an alternative approach and compute the dimuon production cross section directly using semi-inclusive DIS (SIDIS) for the production of charmed hadrons and a decay function based on experimental data to implement the subsequent semimuonic decay of the hadrons. In our first implementation in ref.~\cite{Helenius:2024fow}, the SIDIS cross section was implemented in approximative mass scheme. In the present work \cite{Paukkunen:2025kjb}, we extend the SIDIS calculation to the full simplified Aivazis-Collins-Olness-Tung (SACOT) scheme \cite{Olness:1987ep,Aivazis:1993kh,Aivazis:1993pi,Kretzer:1997pd,Collins:1998rz,Kramer:2000hn} --- specifically, the SACOT-$\chi$ scheme \cite{Tung:2001mv,Kretzer:2003it,Guzzi:2011ew,Gao:2021fle,Risse:2025smp}.

\section{Dimuon production}

In our approach to dimuon production \cite{Helenius:2024fow,Paukkunen:2025kjb}, the full dimuon cross section can be written as
\begin{equation}
    \frac{\dd\sigma(\nu N\to \mu\mu X)}{\dd x \, \dd y}=\sum_h\int \dd z \frac{\dd\sigma(\nu N\to \mu hX)}{\dd x \, \dd y \, \dd z} B_{h\to\mu}(E_h=zyE_\nu, E_\mu^{\text{min}}),
\end{equation}
where the hadron-production process $\nu_\mu(k)+N(P_N)\to \mu(k')+h(P_h)+X$ is described by the SIDIS cross section
\begin{equation}
\label{eq:sidis_hadron_cross_section}
\begin{split}
	\frac{\dd\sigma(\nu_\mu N\to \mu h X)}{\dd x \, \dd y \, \dd z}= \frac{G_F^2M_W^4}{\left(Q^2+M_W^2\right)^2}
  \frac{Q^2}{2\pi xy}\bigg[&xy^2 F_1(x, z, Q^2)+\left(1-y-\frac{xy M^2}{s-M^2}\right)F_2(x, z, Q^2) \\ & \pm xy\left(1-\frac{y}{2}\right)F_3(x, z, Q^2)\bigg],
\end{split}
\end{equation}
with the usual kinematical variables
\begin{equation}
\label{eq:sidis_kinematics}
\begin{aligned}
	Q^2&=-q^2 = -(k-k')^2, \quad x=\frac{Q^2}{2P_N\cdot q}, \quad y=\frac{P_N\cdot q}{P_N\cdot k}, \quad z=\frac{P_N\cdot P_h}{P_N\cdot q}.
\end{aligned}
\end{equation}
The sign of the $F_3$ term depends on the process, and is $+$ for a neutrino beam and $-$ for an antineutrino beam.

The semimuonic decay of the charmed hadron $h$ is described by the energy-dependent branching ratio $B_{h\to \mu}$, which depends on the hadron energy and also the experimentally-imposed energy cut $E_\mu^{\text{min}}$ on the decay muon. The energy-dependent branching fraction is a nonperturbative object, and in our approach is described by a decay function that can be parametrized and fitted to independent experimental data. This approach is described in more details in ref.~\cite{Helenius:2024fow}.

\section{SIDIS in the SACOT-$\chi$ scheme}

Below the charm-mass threshold $Q^2<m_c^2$, the fixed-order structure functions can be written as \cite{Gluck:1997sj}
\begin{equation}
\label{eq:sidis_fixed_order}
    F_i=\sum_q \kappa_i \left|V_{qc}\right|^2q(\fact^2)\otimes \left[1+\frac{\alpha_s}{2\pi}H_i^{qq}\right]\otimes D_{c\to h}+\kappa_i g(\fact^2)\otimes \left[\frac{\alpha_s}{2\pi}H_i^{qg}\right]\otimes D_{c\to h},
\end{equation}
where $V_{qc}$ are the CKM matrix elements, and the coefficient functions $H_i^{qq}$ and $H_i^{qg}$ depend on the factorization scale $\fact^2$ and the charm mass $m_c^2$. The dependence of $\alpha_s$ on the renormalization scale $\renorm^2$ has been omitted from eq.~\eqref{eq:sidis_fixed_order}. In addition, channels involving the fragmentation of a parton other than the charm quark have been omitted from eq.~\eqref{eq:sidis_fixed_order} as they can be regarded as higher-order corrections due to the implicit splitting to a $c\bar c$ pair. These omitted channels are included in the numerical implementation, although their contribution is neglegible \cite{Helenius:2024fow}. The double convolution is given by
\begin{equation}
    q\otimes H\otimes D\equiv\int_\chi^1\frac{\dd\xi}{\xi}\int_{\max(z,\zeta_{\text{min}})}^1\frac{\dd\zeta}{\zeta} q(\chi/\xi)H(\xi, \zeta) D(z/\zeta)
\end{equation}
where
\begin{equation}
    \chi=x\left(1+\frac{m_c^2}{Q^2}\right), \quad \zeta_{\text{min}}=\frac{(1-\lambda)\xi}{1-\lambda\xi}, \quad\text{and}\quad \lambda=\frac{Q^2}{Q^2+m_c^2}.
\end{equation}
The normalizations of the structure functions are given by
\begin{equation}
    \kappa_1 =1, \quad \kappa_2=2\chi, \quad \kappa_3=\pm 2,
\end{equation}
where the sign of $\kappa_3$ is negative in the case of an antineutrino beam.

The quark-to-quark coefficients $H_i^{qq}$ contain logarithmic mass divergences, which must be resummed. This can be achieved by absorbing the divergences to the scale evolution of the fragmentation functions, which are independent of the fragmentation scale $\frag^2$ below the charm threshold. The fragmentation functions, which are evolved to all orders, can be expanded based on the DGLAP equations as
\begin{equation}
    D_{c\to h}(z, \frag^2)=D_{c\to h}(z)+\frac{\alpha_s}{2\pi}\log\left(\frac{\frag^2}{m_c^2}\right)D_{c\to h}\otimes (P_{qq}+\text{scheme})+\mathcal{O}(\alpha_s^2),
\end{equation}
which can be perturbatively inverted as
\begin{equation}
\label{eq:D}
    D_{c\to h}(z)=D_{c\to h}(z, \frag^2)-\frac{\alpha_s}{2\pi}\log\left(\frac{\frag^2}{m_c^2}\right)D_{c\to h}(\frag^2)\otimes (P_{qq}+\text{scheme})+\mathcal{O}(\alpha_s^2),
\end{equation}
where the splitting function is given by the plus distribution
\begin{equation}
    P_{qq}(\zeta)=C_F\left[\frac{1+\zeta^2}{1-\zeta}\right]_+.
\end{equation}
Inserting eq.~\eqref{eq:D} back to eq.~\eqref{eq:sidis_fixed_order} then yields the structure function valid above the charm threshold:
\begin{equation}
\label{eq:sidis_gm}
\begin{aligned}
    F_i&=\sum_q \kappa_i \left|V_{qc}\right|^2q(\fact^2)\otimes \left[1+\frac{\alpha_s}{2\pi}H_i^{qq}\right]\otimes D_{c\to h}(\frag^2)+\kappa_i g(\fact^2)\otimes \left[\frac{\alpha_s}{2\pi}H_i^{qg}\right]\otimes D_{c\to h}(\frag^2) \\
    &\phantom{=} -\frac{\alpha_s}{2\pi}\log\left(\frac{\frag^2}{m_c^2}\right)\sum_q \kappa_i \left|V_{qc}\right|^2 q(\fact^2)\otimes \left[1+\frac{\alpha_s}{2\pi}H_i^{qq}\right]\otimes D_{c\to h}(\frag^2)\otimes (P_{qq}+\text{scheme}).
\end{aligned}
\end{equation}
The second line of eq.~\eqref{eq:sidis_gm} is often referred to as the subtraction term, and it simplifies to
\begin{equation}
    \frac{\alpha_s}{2\pi}q(\chi, \fact^2) (P_{qq}+\text{scheme})\otimes D_{c\to h}(\frag^2),
\end{equation}
where the single convolution is given by
\begin{equation}
    P\otimes D\equiv\int_z^1 \frac{\dd\zeta}{\zeta} P(\zeta)D(z/\zeta).
\end{equation}
The scheme term,
\begin{equation}
    \text{scheme}(\zeta)=-C_F\left[\frac{1+\zeta^2}{1-\zeta}\left(1+2\log(1-\zeta)\right)\right]_+,
\end{equation}
ensures that the usual $\overline{\mathrm{MS}}$ structure functions are recovered in the massless limit $m_c^2/Q^2\to 0$.

\section{Results}

\begin{figure}[ht!]
	\centering
	\begin{subfigure}{0.3\textwidth}
		\includegraphics[width=\linewidth]{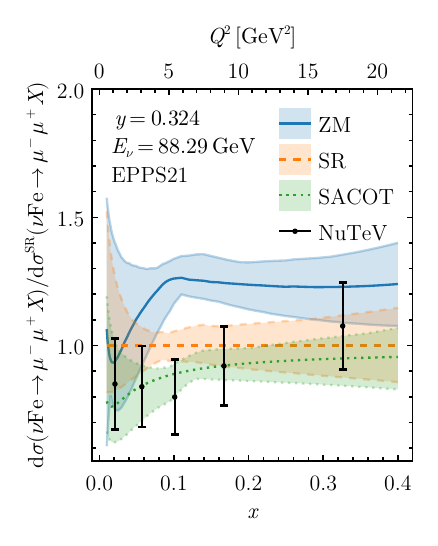}
	\end{subfigure}%
	\begin{subfigure}{0.3\textwidth}
		\includegraphics[width=\linewidth]{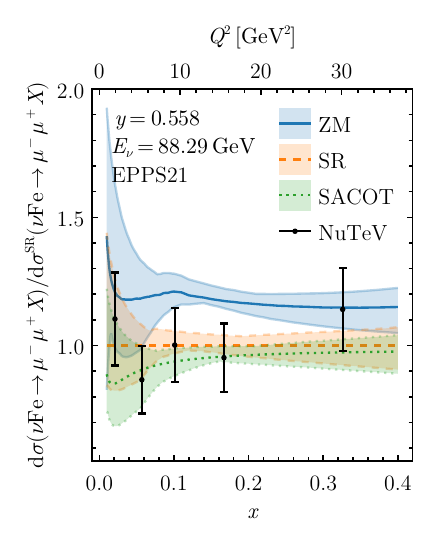}
	\end{subfigure}%
	\begin{subfigure}{0.3\textwidth}
		\includegraphics[width=\linewidth]{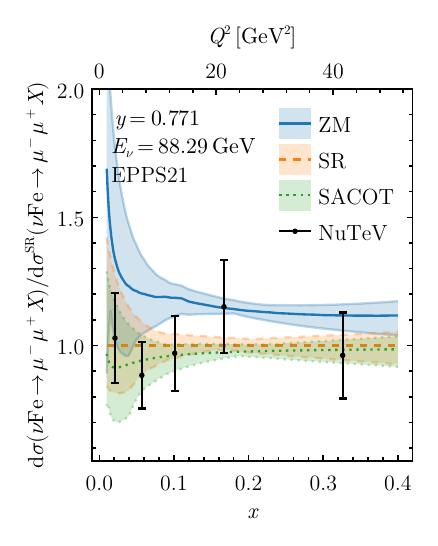}
	\end{subfigure}
	\begin{subfigure}{0.3\textwidth}
		\includegraphics[width=\linewidth]{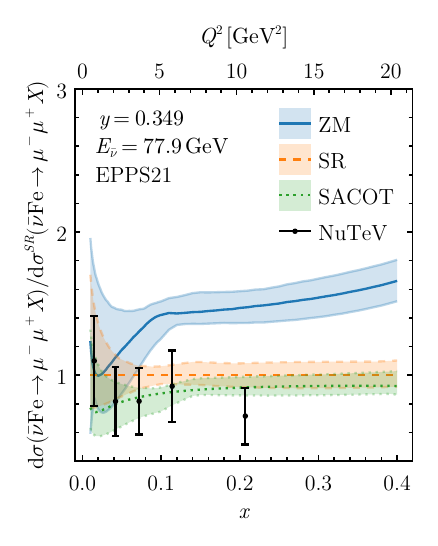}
	\end{subfigure}%
	\begin{subfigure}{0.3\textwidth}
		\includegraphics[width=\linewidth]{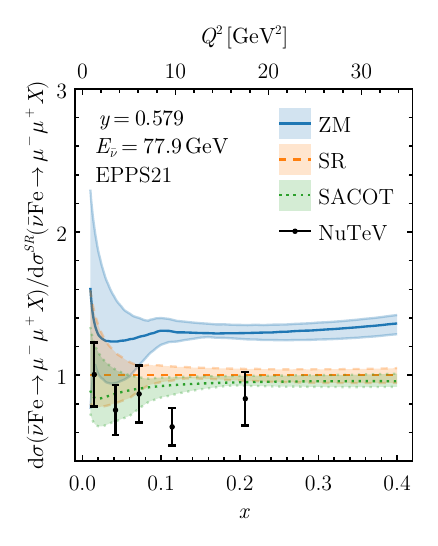}
	\end{subfigure}%
	\begin{subfigure}{0.3\textwidth}
		\includegraphics[width=\linewidth]{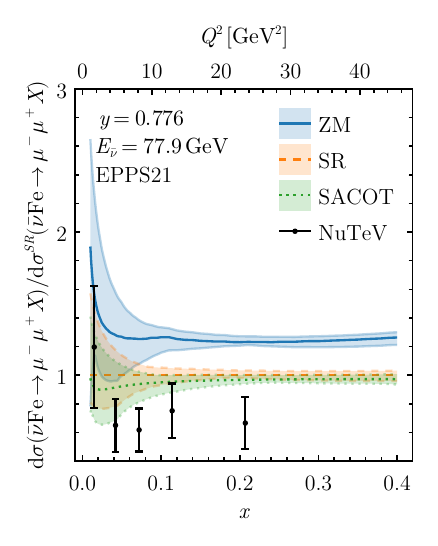}
	\end{subfigure}
	\caption{Comparison of mass schemes with the \texttt{EPPS21} nPDF at NLO with $\mu^2=Q^2+m_c^2$ in neutrino (top row) and antineutrino (bottom row) scattering. The uncertainty band depicts the envelope of all 17 scale combinations. The theoretical calculations and experimental NuTeV data are normalized to the central SR curve.}
	\label{fig:comparison}
\end{figure}

Figure \ref{fig:comparison} presents a comparison of three different mass schemes for dimuon production in (anti)neutrino scattering: the zero-mass (ZM), the slow-rescaling (SR), which is also referred to as the intermediate-mass (IM), and the SACOT-$\chi$ scheme. The ZM scheme does not include any charm-mass effects. The SR scheme only includes kinematical mass effects, and represents our original calculation done in ref.~\cite{Helenius:2024fow}. The SACOT-$\chi$ scheme includes both kinematical and dynamical mass corrections.

The dynamical charm-mass corrections are subleading, as there are no matrix-element level mass corrections at leading order. This is a unique feature of charged-current DIS with one massive quark, which means that the SR scheme is already a good approximation of the full SACOT-$\chi$ calculation. The dynamical mass corrections appearing at NLO are of the form $m_c^2/Q^2$ and therefore most important at small $Q^2$. This is indeed what can be seen in fig.~\ref{fig:comparison}, as the SACOT-$\chi$ scheme calculation approaches the SR calculation as $Q^2$ grows. At small $Q^2$, however, the SACOT-$\chi$ calculation modifies the SR calculation by up to $20 \, \%$.

\section{Summary}

We have presented a SIDIS-based approach to calculating dimuon production in neutrino-nucleus collisions, now in the full SACOT-$\chi$ scheme. Our approach, unlike the usual one, does not use inclusive charm production but instead SIDIS for charmed-hadron production. This gives an approach that can be systematically improved to higher perturbative orders and mass schemes. In this work, we extended our previous calculation to the SACOT-$\chi$ scheme, which takes all charm-mass effects fully and consistently into account. The results are in line with our expectations: the corrections from dynamical mass effects are most notable (up to $20 \, \%$) at small values of $Q^2$. As $Q^2$ grows, the SACOT-$\chi$ calculation approaches the approximative SR scheme.

In the future, we will implement the NNLO corrections (in the approximative SR scheme) using the recently-released NNLO coefficients for charged-current SIDIS \cite{Bonino:2025qta}.

\acknowledgments

We acknowledge the financial support from the Magnus Ehrnrooth foundation (S.Y.), the Research Council of Finland Project No. 331545 and 361179 (I.H.), and the Center of Excellence in Quark Matter of the Research Council of Finland, project 346326. The reported work is associated with the European Research Council project ERC-2018-ADG-835105 YoctoLHC. We acknowledge grants of computer capacity from the Finnish Grid and Cloud Infrastructure (persistent identifier
   \texttt{urn:nbn:fi:research-infras-2016072533}) and the Finnish IT Center for Science (CSC), under the project jyy2580.

\bibliographystyle{JHEP}
\bibliography{Refs.bib}

\providecommand{\href}[2]{#2}\begingroup\raggedright\begin{thebibliography}{10}

\bibitem{Hou:2019efy}
T.-J.~Hou et~al., \emph{{New CTEQ global analysis of quantum chromodynamics with high-precision data from the LHC}}, \href{https://doi.org/10.1103/PhysRevD.103.014013}{\emph{Phys. Rev. D} {\bfseries 103} (2021) 014013} [\href{https://arxiv.org/abs/1912.10053}{{\ttfamily 1912.10053}}].

\bibitem{Bailey:2020ooq}
S.~Bailey, T.~Cridge, L.A.~Harland-Lang, A.D.~Martin and R.S.~Thorne, \emph{{Parton distributions from LHC, HERA, Tevatron and fixed target data: MSHT20 PDFs}}, \href{https://doi.org/10.1140/epjc/s10052-021-09057-0}{\emph{Eur. Phys. J. C} {\bfseries 81} (2021) 341} [\href{https://arxiv.org/abs/2012.04684}{{\ttfamily 2012.04684}}].

\bibitem{NNPDF:2021njg}
{\scshape NNPDF} collaboration, \emph{{The path to proton structure at 1\% accuracy}}, \href{https://doi.org/10.1140/epjc/s10052-022-10328-7}{\emph{Eur. Phys. J. C} {\bfseries 82} (2022) 428} [\href{https://arxiv.org/abs/2109.02653}{{\ttfamily 2109.02653}}].

\bibitem{Eskola:2021nhw}
K.J.~Eskola, P.~Paakkinen, H.~Paukkunen and C.A.~Salgado, \emph{{EPPS21: a global QCD analysis of nuclear PDFs}}, \href{https://doi.org/10.1140/epjc/s10052-022-10359-0}{\emph{Eur. Phys. J. C} {\bfseries 82} (2022) 413} [\href{https://arxiv.org/abs/2112.12462}{{\ttfamily 2112.12462}}].

\bibitem{Duwentaster:2022kpv}
P.~Duwent\"aster, T.~Je\v{z}o, M.~Klasen, K.~Kova\v{r}\'\i{}k, A.~Kusina, K.F.~Muzakka et~al., \emph{{Impact of heavy quark and quarkonium data on nuclear gluon PDFs}}, \href{https://doi.org/10.1103/PhysRevD.105.114043}{\emph{Phys. Rev. D} {\bfseries 105} (2022) 114043} [\href{https://arxiv.org/abs/2204.09982}{{\ttfamily 2204.09982}}].

\bibitem{AbdulKhalek:2022fyi}
R.~Abdul~Khalek, R.~Gauld, T.~Giani, E.R.~Nocera, T.R.~Rabemananjara and J.~Rojo, \emph{{nNNPDF3.0: evidence for a modified partonic structure in heavy nuclei}}, \href{https://doi.org/10.1140/epjc/s10052-022-10417-7}{\emph{Eur. Phys. J. C} {\bfseries 82} (2022) 507} [\href{https://arxiv.org/abs/2201.12363}{{\ttfamily 2201.12363}}].

\bibitem{CCFR:1994ikl}
{\scshape CCFR} collaboration, \emph{{Determination of the strange quark content of the nucleon from a next-to-leading order QCD analysis of neutrino charm production}}, \href{https://doi.org/10.1007/BF01571875}{\emph{Z. Phys. C} {\bfseries 65} (1995) 189} [\href{https://arxiv.org/abs/hep-ex/9406007}{{\ttfamily hep-ex/9406007}}].

\bibitem{NuTeV:2001dfo}
{\scshape NuTeV} collaboration, \emph{{Precise Measurement of Dimuon Production Cross-Sections in $\nu_{\mu}$ Fe and $\bar{\nu}_{\mu}$ Fe Deep Inelastic Scattering at the Tevatron.}}, \href{https://doi.org/10.1103/PhysRevD.64.112006}{\emph{Phys. Rev. D} {\bfseries 64} (2001) 112006} [\href{https://arxiv.org/abs/hep-ex/0102049}{{\ttfamily hep-ex/0102049}}].

\bibitem{NuTeV:2007uwm}
{\scshape NuTeV} collaboration, \emph{{Measurement of the Nucleon Strange-Antistrange Asymmetry at Next-to-Leading Order in QCD from NuTeV Dimuon Data}}, \href{https://doi.org/10.1103/PhysRevLett.99.192001}{\emph{Phys. Rev. Lett.} {\bfseries 99} (2007) 192001}.

\bibitem{CHORUS:2008vjb}
{\scshape CHORUS} collaboration, \emph{{Leading order analysis of neutrino induced dimuon events in the CHORUS experiment}}, \href{https://doi.org/10.1016/j.nuclphysb.2008.02.013}{\emph{Nucl. Phys. B} {\bfseries 798} (2008) 1} [\href{https://arxiv.org/abs/0804.1869}{{\ttfamily 0804.1869}}].

\bibitem{NOMAD:2013hbk}
{\scshape NOMAD} collaboration, \emph{{A Precision Measurement of Charm Dimuon Production in Neutrino Interactions from the NOMAD Experiment}}, \href{https://doi.org/10.1016/j.nuclphysb.2013.08.021}{\emph{Nucl. Phys. B} {\bfseries 876} (2013) 339} [\href{https://arxiv.org/abs/1308.4750}{{\ttfamily 1308.4750}}].

\bibitem{Helenius:2024fow}
I.~Helenius, H.~Paukkunen and S.~Yrj\"anheikki, \emph{{Dimuons from neutrino-nucleus collisions in the semi-inclusive DIS approach}}, \href{https://doi.org/10.1007/JHEP09(2024)043}{\emph{JHEP} {\bfseries 09} (2024) 043} [\href{https://arxiv.org/abs/2405.12677}{{\ttfamily 2405.12677}}].

\bibitem{Paukkunen:2025kjb}
H.~Paukkunen, I.~Helenius and S.~Yrj{\"a}nheikki, \emph{{Improving the description of dimuon production in neutrino-nucleus collisions using the SACOT-$\chi$ scheme}},  \href{https://arxiv.org/abs/2506.09492}{{\ttfamily 2506.09492}}.

\bibitem{Olness:1987ep}
F.I.~Olness and W.-K.~Tung, \emph{{When Is a Heavy Quark Not a Parton? Charged Higgs Production and Heavy Quark Mass Effects in the QCD Based Parton Model}}, \href{https://doi.org/10.1016/0550-3213(88)90129-0}{\emph{Nucl. Phys. B} {\bfseries 308} (1988) 813}.

\bibitem{Aivazis:1993kh}
M.A.G.~Aivazis, F.I.~Olness and W.-K.~Tung, \emph{{Leptoproduction of heavy quarks. 1. General formalism and kinematics of charged current and neutral current production processes}}, \href{https://doi.org/10.1103/PhysRevD.50.3085}{\emph{Phys. Rev. D} {\bfseries 50} (1994) 3085} [\href{https://arxiv.org/abs/hep-ph/9312318}{{\ttfamily hep-ph/9312318}}].

\bibitem{Aivazis:1993pi}
M.A.G.~Aivazis, J.C.~Collins, F.I.~Olness and W.-K.~Tung, \emph{{Leptoproduction of heavy quarks. 2. A Unified QCD formulation of charged and neutral current processes from fixed target to collider energies}}, \href{https://doi.org/10.1103/PhysRevD.50.3102}{\emph{Phys. Rev. D} {\bfseries 50} (1994) 3102} [\href{https://arxiv.org/abs/hep-ph/9312319}{{\ttfamily hep-ph/9312319}}].

\bibitem{Kretzer:1997pd}
S.~Kretzer and I.~Schienbein, \emph{{Charged current leptoproduction of D mesons in the variable flavor scheme}}, \href{https://doi.org/10.1103/PhysRevD.56.1804}{\emph{Phys. Rev. D} {\bfseries 56} (1997) 1804} [\href{https://arxiv.org/abs/hep-ph/9702296}{{\ttfamily hep-ph/9702296}}].

\bibitem{Collins:1998rz}
J.C.~Collins, \emph{{Hard scattering factorization with heavy quarks: A General treatment}}, \href{https://doi.org/10.1103/PhysRevD.58.094002}{\emph{Phys. Rev. D} {\bfseries 58} (1998) 094002} [\href{https://arxiv.org/abs/hep-ph/9806259}{{\ttfamily hep-ph/9806259}}].

\bibitem{Kramer:2000hn}
M.~Kr\"amer, F.I.~Olness and D.E.~Soper, \emph{{Treatment of heavy quarks in deeply inelastic scattering}}, \href{https://doi.org/10.1103/PhysRevD.62.096007}{\emph{Phys. Rev. D} {\bfseries 62} (2000) 096007} [\href{https://arxiv.org/abs/hep-ph/0003035}{{\ttfamily hep-ph/0003035}}].

\bibitem{Tung:2001mv}
W.-K.~Tung, S.~Kretzer and C.~Schmidt, \emph{{Open heavy flavor production in QCD: Conceptual framework and implementation issues}}, \href{https://doi.org/10.1088/0954-3899/28/5/321}{\emph{J. Phys. G} {\bfseries 28} (2002) 983} [\href{https://arxiv.org/abs/hep-ph/0110247}{{\ttfamily hep-ph/0110247}}].

\bibitem{Kretzer:2003it}
S.~Kretzer, H.L.~Lai, F.I.~Olness and W.K.~Tung, \emph{{Cteq6 parton distributions with heavy quark mass effects}}, \href{https://doi.org/10.1103/PhysRevD.69.114005}{\emph{Phys. Rev. D} {\bfseries 69} (2004) 114005} [\href{https://arxiv.org/abs/hep-ph/0307022}{{\ttfamily hep-ph/0307022}}].

\bibitem{Guzzi:2011ew}
M.~Guzzi, P.M.~Nadolsky, H.-L.~Lai and C.P.~Yuan, \emph{{General-Mass Treatment for Deep Inelastic Scattering at Two-Loop Accuracy}}, \href{https://doi.org/10.1103/PhysRevD.86.053005}{\emph{Phys. Rev. D} {\bfseries 86} (2012) 053005} [\href{https://arxiv.org/abs/1108.5112}{{\ttfamily 1108.5112}}].

\bibitem{Gao:2021fle}
J.~Gao, T.J.~Hobbs, P.M.~Nadolsky, C.~Sun and C.P.~Yuan, \emph{{General heavy-flavor mass scheme for charged-current DIS at NNLO and beyond}}, \href{https://doi.org/10.1103/PhysRevD.105.L011503}{\emph{Phys. Rev. D} {\bfseries 105} (2022) L011503} [\href{https://arxiv.org/abs/2107.00460}{{\ttfamily 2107.00460}}].

\bibitem{Risse:2025smp}
P.~Risse, V.~Bertone, T.~Je\v{z}o, K.~Kova\v{r}\'\i{}k, A.~Kusina, F.~Olness et~al., \emph{{Heavy Quark mass effects in charged-current Deep-Inelastic Scattering at NNLO in the ACOT scheme}},  \href{https://arxiv.org/abs/2504.13317}{{\ttfamily 2504.13317}}.

\bibitem{Gluck:1997sj}
M.~Gluck, S.~Kretzer and E.~Reya, \emph{{Detailed next-to-leading order analysis of deep inelastic neutrino induced charm production off strange sea partons}}, \href{https://doi.org/10.1016/S0370-2693(97)00232-3}{\emph{Phys. Lett. B} {\bfseries 398} (1997) 381} [\href{https://arxiv.org/abs/hep-ph/9701364}{{\ttfamily hep-ph/9701364}}].

\bibitem{Bonino:2025qta}
L.~Bonino, T.~Gehrmann, M.~L{\"o}chner, K.~Sch{\"o}nwald and G.~Stagnitto, \emph{{Neutral and charged current semi-inclusive deep-inelastic scattering at NNLO QCD}}, \href{https://doi.org/10.1007/JHEP10(2025)016}{\emph{JHEP} {\bfseries 10} (2025) 016} [\href{https://arxiv.org/abs/2506.19926}{{\ttfamily 2506.19926}}].

\end{thebibliography}\endgroup

\end{document}